\documentclass[conference]{IEEEtran} 

\usepackage{bm}
\usepackage[pdftex]{graphicx}
\DeclareGraphicsExtensions{.pdf,.jpeg,.png}
\usepackage{epstopdf}
\usepackage{tikz}
\usetikzlibrary{patterns}
\usepackage{pgfplots}
\usepackage{amsmath}
\usepackage{dsfont}
\allowdisplaybreaks[4]

\usepackage{cite}

\ifCLASSINFOpdf
\else
\fi
\usepackage{amsmath}
\usepackage{algorithmic}

\usepackage{array}

\ifCLASSOPTIONcompsoc
  \usepackage[caption=false,font=normalsize,labelfont=sf,textfont=sf]{subfig}
\else
  \usepackage[caption=false,font=footnotesize]{subfig}
\fi
\usepackage{url}
\usepackage{mathalfa}
\usepackage{amsfonts}
\usepackage{color}
\usepackage{makecell}

\graphicspath{{fig/}}

\begin{document}

%
\title{Secure Visible Light Communications via Intelligent Reflecting Surfaces}
%
%
%
\author{\IEEEauthorblockN{Lei Qian\textsuperscript{1}  , Xuefen Chi\textsuperscript{1,*}  , Linlin Zhao\textsuperscript{1}  , Anas Chaaban\textsuperscript{2}  }
\IEEEauthorblockA{\textsuperscript{1}  Department of Communications Engineering, Jilin University, Changchun, China\\
\textsuperscript{2}  School of Engineering, University of British Columbia, Kelowna, BC, Canada\\
 Email: qianlei16@mails.jlu.edu.cn, chixf@jlu.edu.cn, zhaoll13@mails.jlu.edu.cn, anas.chaaban@ubc.ca }}
\IEEEaftertitletext{\vspace{-1.5\baselineskip}}
\maketitle
\begin{abstract}
Intelligent reflecting surfaces (IRS) can improve the physical layer security (PLS) by providing a controllable wireless environment. 
In this paper, we propose a novel PLS technique with the help of IRS implemented by an intelligent mirror array for the visible light communication (VLC) system.
First, for the IRS aided VLC system containing an access point (AP), a legitimate user and an eavesdropper, the IRS channel gain and a lower bound of the achievable secrecy rate are derived.
Further, to enhance the IRS channel gain of the legitimate user while restricting the IRS channel gain of the eavesdropper, we formulate an achievable secrecy rate maximization problem for the proposed IRS-aided PLS technique to find the optimal orientations of mirrors.
Since the sensitivity of mirrors' orientations on the IRS channel gain makes the optimization problem hard to solve, we transform the original problem into a reflected spot position optimization problem and solve it by a particle swarm optimization (PSO) algorithm.
Our simulation results show that secrecy performance can be significantly improved by adding an IRS in a VLC system. 
\end{abstract}


%
\IEEEpeerreviewmaketitle

\section{Introduction}
%
%
%
%
\IEEEPARstart{S}{ecurity} and privacy guarantees attract special attention in the sixth generation (6G) networks \cite{6G-Shanzhi} \cite{6G-Nature}. 
Inherited from the fact that light cannot penetrate through walls or opaque objects, visible light communication (VLC) is considered as a secure alternative for the 6G indoor wireless networks \cite{6G-Shanzhi}. 
Furthermore, by employing light-emitting diodes (LEDs), VLC can provide both illumination and high-rate data communication simultaneously.
However, due to the open and broadcasting nature of VLC, user's confidential information may still be intercepted when VLC systems are deployed in public areas, such as shopping centers, libraries, coffee shops, hospitals, train stations, etc. 
During the past few years, physical-layer security (PLS) techniques are introduced to VLC networks in an effort to enhance the overall system security by complementing existing upper-layer encryption security techniques.
For the single-input-single-output (SISO) VLC system, upper and lower bounds on secrecy capacity are derived in \cite{Lampe-2015-JSAC}-\hspace{1sp}\cite{Charlus}. 
Based on Polar Codes, a physical-layer secure coding scheme is proposed to enhance the secrecy performance of SISO VLC system \cite{PLS-polar-codes}.
For a multiple-input-single-output (MISO) VLC system, beamforming \cite{Lampe-2015-JSAC}\cite{Oxford-2018-TWC}\cite{Mashuai} and jamming \cite{Lampe-2014-globecom}-\hspace{1sp}\cite{Anas-2019-TCOM} are two main categories of PLS techniques. 
More specifically, beamforming schemes perform better than jamming schemes when the eavesdropper is an authorized user and the CSI of eavesdropping channel is perfectly known.
By contrast, jamming schemes are most commonly adopted for the cases when the CSI of the eavesdroppers is not available, for example, when eavesdroppers are malicious users not registered in the network \cite{Lampe-2014-globecom}. 
However, both beamforming and jamming techniques require the cooperation of multiple transmitters, which means energy consumption is increased to enhance the secrecy performance of VLC systems.

Recently, as an eye-catching architecture for 6G networks, intelligent reflecting surfaces (IRS) have received significant research attention in radio frequency (RF) systems \cite{6G-Nature}. 
Consisting a number of low-cost passive reflecting elements, an IRS can reconfigure the wireless propagation environment via intelligently controlled reflection \cite{survey-IRS}. 
By adjusting the reflecting coefficients of reflecting elements, IRS can achieve fine-grained three-dimensional (3D) passive beamforming toward the intended receiver.  
Based on the idea of passive beamforming, a new PLS technique is proposed for the RF systems with existence of eavesdroppers \cite{secure-IRS-WCL2019}-\hspace{1sp}\cite{secure-IRS-Access2019}.  
It has been proved that secrecy performance can be significantly improved with the help of IRS by jointly optimizing the reflecting coefficients.
In the same spirit, the IRS can also provide new horizon for the design of PLS technique in VLC secure systems.
 
In VLC systems, IRS can be implemented by programmable metasurfaces or by mirror arrays \cite{KAUST-IRSVLC}.
The phase gradient of each metasurface and the orientation of each mirror can be controlled intelligently to reflect the incident optical signal towards a receiver.
Recently, the irradiance (power density) expressions of two types of reflectors are derived in \cite{KAUST-IRSVLC} to characterize the focusing capability of IRS in the VLC system. 
According to the results in \cite{KAUST-IRSVLC},  the mirror array outperforms the metasurface in VLC system. 
Inspired by \cite{KAUST-IRSVLC}, in this paper, we propose an IRS aided secure VLC system to explore the secrecy improvement brought by IRS in VLC.
The main contributions of this paper are summarized as follows.

\begin{itemize}
\item Using an intelligent controllable mirror array as  IRS, we model the reflected channel gain in a VLC system. We derive a lower bound on the achievable secrecy rate for the IRS-aided peak-constrained VLC system employing intensity-modulation/direct detection (IM/DD).  
\item To enhance secrecy performance of the SISO VLC system, we use an optimized intelligent rotation scheme for the mirrors to enlarge the difference between the channel gain of the legitimate user and the eavesdropper.
Hence, we treat the orientation of each mirror as the optimization variable and formulate an achievable secrecy rate maximization problem for the proposed IRS-based PLS technique.
\item Due to the sensitivity of mirrors' orientations to the IRS channel gain, it is hard to find the optimal combination of mirrors' orientations. To handle this problem, we transform the orientations optimization problem into a reflected spot position finding problem. 
Since the maximum IRS channel gain is achieved when the reflected spots of different mirrors focus on the same position, we just need to find only one position of the reflected spot for the mirror array, which greatly reduce the complexity of the problem.

\end{itemize}

The rest of this paper is organized as follows. The system model is introduced in Sec. \ref{section-system model}. The problem formulation and the solution are introduced in Sec. \ref{section-problem formulation}. Simulation results and discussion are presented and discussed in Sec. \ref{section-simulation}. Finally, the paper is concluded in Sec. \ref{section-conclusion}.
 
\section{System Model}
\label{section-system model}
In this paper, we consider a typical indoor space, whose size is $x_r \times y_r \times z_r \; (\rm{m} \times \rm{m} \times\rm{m}) $, as shown in Fig. 1. 
An extended planar source having uniform
radiant emittance over its area $w_s \times l_s$ is deployed on the ceiling of the room and considered as an optical access point (AP).
The system includes two users: a legitimate user (i.e. Bob) and an eavesdropper (i.e. Eve). 
A mirror array consisting of $N_m \times N_n$ identical rectangular mirrors is installed on the one wall. 
The size of each mirror is $w_m \times h_m$. 
The orientation of each mirror can be rotated independently. 
The rotation angle of the mirror ($i,j$) includes yaw angle $\beta_{i,j}$ and roll angle $\alpha_{i,j}$, as shown in Fig. 2.

\begin{figure}[!t]

\centering
\includegraphics[width=0.47\textwidth]{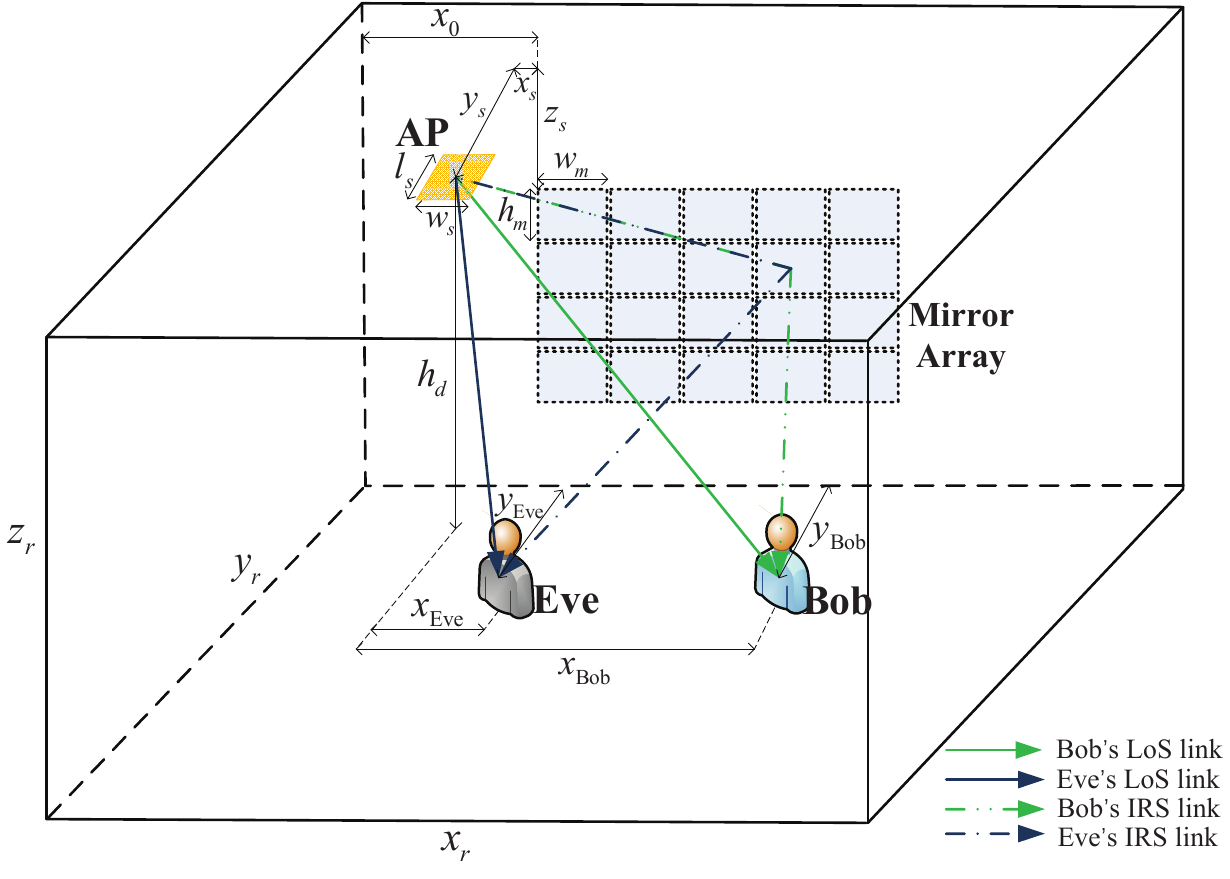}%
\vspace*{-4mm}
\caption{Mirror array aided indoor VLC IRS system.}
\label{Fig-scenario}
\vspace*{-4mm}
\end{figure}

\begin{figure}[!t]
\subfloat[\label{rotation angle yaw}]{%
\centering
\includegraphics[width=0.2\textwidth]{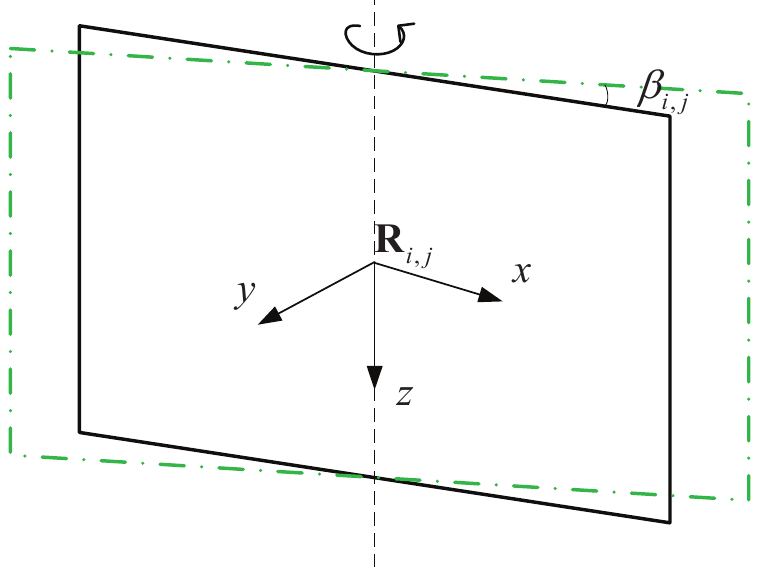}%
}
\hfill
  \subfloat[\label{rotation angle pitch}]{%
  \centering
\includegraphics[width=0.23\textwidth]{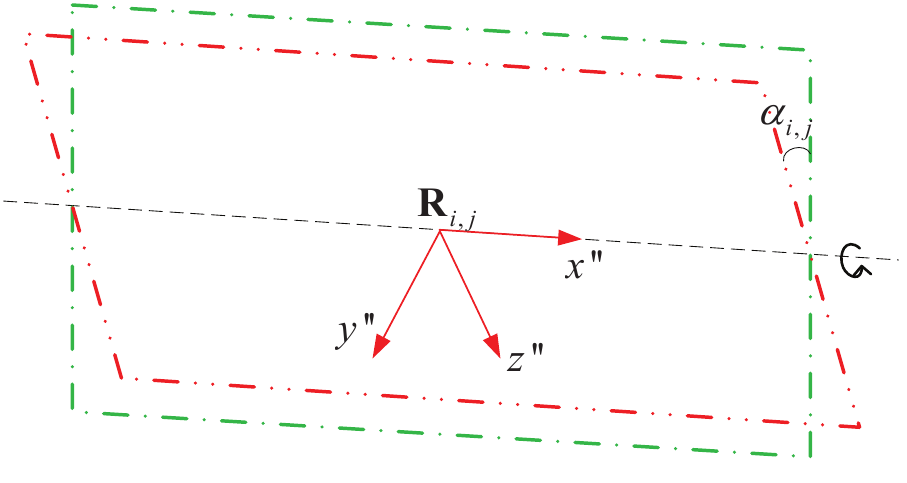}%
}
\caption{(a) The yaw rotation angle of the mirror in $i$th row and $j$th column. (b) The pitch rotation angle of the mirror in $i$th row and $j$th column.}
\vspace*{-4mm}
\end{figure}

A direct current (DC)-biased IM/DD scheme is considered, in which the LED is driven by a fixed bias $I_{DC}$. 
The optical transmitted signal $s \in \mathbb{R}$ is a zero-mean current signal superimposed on $I_{DC}$ to modulate the instantaneous optical power emitted from the LED, where $\mathbb{R}$ represents the set of real numbers. 
To avoid clipping distortion, the signal $s$ needs to satisfy the peak-power constraint characterized by $|s| \le A$, where $A$ is a constant \cite{Lampe-2015-JSAC}.

Due to the broadcast nature of visible light, Eve may wiretap the confidential message of Bob transmitted by AP via both line-of-sight (LoS) links and IRS links\footnote{Since the channel gain of diffused light links are so light compared to the LoS links, the diffused light links are not considered in this paper \cite{2004-NLOS}. }. 
Therefore, the received signal of Bob and the wiretap signal of Eve are expressed as
\begin{equation}
\label{eq-signal-Bob}
 {y^{{\rm{Bob}}}} = ({h^{{\rm{LoS,}}\;{\rm{Bob}}}} + {h^{{\rm{IRS,}}\;{\rm{Bob}}}})s + {n^{{\rm{Bob}}}},
\end{equation}
and
 \begin{equation}
 \label{eq-signal-Eve}
 {y^{{\rm{Eve}}}} = ({h^{{\rm{LoS,}}\;{\rm{Eve}}}} + {h^{{\rm{IRS,}}\;{\rm{Eve}}}})s + {n^{{\rm{Eve}}}}, 
\end{equation}
respectively, where ${n^{{\rm{Bob}}}}\sim{{\mathcal N}(0,\sigma^2)}$ and ${n^{{\rm{Eve}}}}\sim{{\mathcal N}(0,\sigma^2)}$ represent the VLC noise of Bob and Eve, which contains shot noise and thermal noise. The VLC noise is regarded as additive white Gaussian noise with zero-mean and variance $\sigma^2=N_0 B$, where $N_0 \approx 10^{-22} \;\rm{A^2/Hz}$ \cite{2008-VLCnoise}. In addition, ${h^{{\rm{LoS,}}\;{\rm{Bob}}}}$ and ${h^{{\rm{LoS,}}\;{\rm{Eve}}}}$ represent the DC channel gain of LoS links of Bob and Eve, while ${h^{{\rm{IRS,}}\;{\rm{Bob}}}}$ and ${h^{{\rm{IRS,}}\;{\rm{Eve}}}}$ represent the IRS channel gain of LoS links of Bob and Eve.

For a LoS link, the channel gain is expressed as \cite{2004-NLOS}
\begin{equation}
\label{eq-LoS-gain}
\begin{aligned} 
{h^{{\rm{LoS}},\:k}}{\rm{ = }}\int_{ - {{{w_s}} \mathord{\left/
 {\vphantom {{{w_s}} 2}} \right.
 \kern-\nulldelimiterspace} 2}}^{{{{w_s}} \mathord{\left/
 {\vphantom {{{w_s}} 2}} \right.
 \kern-\nulldelimiterspace} 2}} {\int_{{{ - {l_s}} \mathord{\left/
 {\vphantom {{ - {l_s}} 2}} \right.
 \kern-\nulldelimiterspace} 2}}^{{{{l_s}} \mathord{\left/
 {\vphantom {{{l_s}} 2}} \right.
 \kern-\nulldelimiterspace} 2}} {\frac{{\eta ({L_a} + 1)\delta \varpi T}}{{2\pi d_k^2(x,y)}}g(\psi _k^{{\rm{LoS}},{\rm{in}}}(x,y))}} \\
 {{\cos }^{{L_a}}}(\psi _k^{{\rm{LoS}},{\rm{ir}}}(x,y))\cos (\psi _k^{{\rm{LoS}},{\rm{in}}}(x,y))dxdy  ,
\end{aligned} 
\end{equation}
where $k \in \{ {\rm{Bob}},{\rm{Eve}}\}$, $d_k(x,y)$ is the distance between AP and the user $k$, $L_a =  - \ln 2/\ln(\cos({\phi _{1/2}}))$ denotes the order of Lambertian emission, $\phi_{1/2}$ is the semi-angle at half illumination of LEDs, $\delta$ is the physical area of the detector in a PD, $\varpi$ is responsivity of a PD, $\eta$ is the current-to-light conversion efficiency of
the LED, $T$ is the gain of the transimpedance amplifier, $\psi _{k}^{{\rm{LoS,in}}}(x,y)$, and $\psi _{k}^{{\rm{LoS,ir}}}(x,y)$ are the angle of incidence and irradiance between the AP and the user $k$, respectively. In addition, the function $g(\cdot)$ represents the gain of an optical concentrator, which is characterized by 
\begin{equation}
\label{optical-concentrater}
g(\psi _{k}^{{\rm{LoS}},{\rm{in}}}(x,y)) = \left\{ \begin{array}{l}
{{{a^2}} \mathord{\left/
 {\vphantom {{{a^2}} {{{\sin }^2}{\varphi_{\rm{c}}}}}} \right.
 \kern-\nulldelimiterspace} {{{\sin }^2}{\varphi_{\rm{c}}}}},\;{\rm{if}}\;\psi _{k}^{{\rm{LoS}},{\rm{in}}}(x,y) \le {\varphi _{\rm{c}}}, \\
 [1mm]
0,\;\;\;\;\;\;\;\;\;\;\;\;\;\;{\rm{if}}\;\psi _{k}^{{\rm{LoS}},{\rm{in}}}(x,y) > {\varphi _{\rm{c}}},
\end{array} \right.
\end{equation}
where $a$ denotes the refractive index, and $\varphi_{\rm{c}}$ is half of FoV at a receiver.

In this paper, our objective is using an optimized intelligent rotation scheme for the mirrors to improve the secrecy performance of the VLC wiretap channel in (\ref{eq-signal-Bob}) and (\ref{eq-signal-Eve}).
\section{Problem Formulation and Solution}
\label{section-problem formulation}
\subsection{IRS Channel gain}
\label{subsection-IRS channel gain}

For a VLC IRS system implemented by an intelligent mirror array, the IRS channel gain of user $k$ is derived as 
\begin{equation}
\label{eq-IRS-gain}
h^{{\rm{IRS}},\:k}({\bm{\alpha }},{\bm{\beta }}) = \sum\limits_{i = 1,j = 1}^{\rm{all\;mirrors}} {\eta \varpi \delta TE_{i,j}^k({\bm{\alpha }},{\bm{\beta }})g(\theta _{{\mathbf{R}}_{i,j}}^{{\mathbf{P}}_k})\cos (\theta _{{\mathbf{R}}_{i,j}}^{{\mathbf{P}}_k})}, 
\end{equation} 
where $E_{i,j}^{k}({\bm{\alpha }},{\bm{\beta }})$ represents the irradiance at user $k$ contributed by the mirror ($i,j$). 
Based on geometrical optics and trigonometric analysis, $E_{i,j}^{k}({\bm{\alpha }},{\bm{\beta }})$ is expressed as \cite{KAUST-IRSVLC}
\begin{equation}
\label{eq-IRS-irradience}
\begin{array}{l}
 E_{i,j}^{k}({\bm{\alpha }},{\bm{\beta }}) 
  \vspace{1ex} \\
 = \frac{{(L_a + 1)\rho}}{{2\pi }}\int_{ - {h_m}/2}^{{h_m}/2} \int_{ - {w_m}/2}^{{w_m}/2} {{\cos }^{(L_a)}}(\theta _{{\bf{R}}_{i,j}}^{\bf{I}})
 \vspace{1ex} \\
 \times  \frac{{\mathbf{e}_3^T({\bf{P}}_k - {{\bf{R}}_{i,j}}){\widehat{\mathbf{N}}_{i,j}}^T({\bf{P}}_k - {{\bf{R}}_{i,j}})}}{{\left\| {{\bf{P}}_k - {{\bf{R}}_{i,j}}} \right\|_2^4}}    
 \vspace{1ex} \\ 
 \times \mathbb{I} ( \mathbf{e}_1^T{\bf{S}} - \frac{{{w_s}}}{2} \le \mathbf{e}_1^T{\bf{I}} \le \mathbf{e}_1^T{\bf{S}} + \frac{{{w_s}}}{2}, 
  \vspace{1ex} \\
\mathbf{e}_2^T{\bf{S}} - \frac{{{l_s}}}{2} \le \mathbf{e}_2^T{\bf{I}} \le \mathbf{e}_2^T{\bf{S}} + \frac{{{l_s}}}{2} ) dx''dz'', \\ 
 \end{array}
\end{equation}
where $\rho$ denotes the reflection efficiency, $\mathbf{e}_n (n \in \{1,2,3\})$ denotes the $n$th column in the $3\times3$ identity matrix, ${\left\| \cdot \right\|}_2$ denotes the $l_2$-norm, and $\mathbb{I}(\cdot)$ denotes the binary indicator function. In addition, $\bf{S}$, ${\bf{P}}_k$, ${{\bf{R}}_{i,j}}$ and ${\bf{I}}$ represent the space coordinates of the center of AP, the position of the user $k$, the differential point of the mirror in the $i$th row $j$th column (i.e. the mirror ($i,j$)), and the pre-reflection image of ${\bf{P}}_k$ in the source plane, respectively. 
According to \cite{KAUST-IRSVLC}, the local Cartesian coordinate system is defined with the origin at the mirror center and the above space coordinates are given as
\begin{equation}
\begin{gathered}
  {{\mathbf{R}}_{i,j}} =\mathbb{R}_{i,j}^{r}{\left[ {x''{\text{  }}0{\text{  }}z''} \right]^T}
, \hfill \\
  {\mathbf{S}} = \left[ {\begin{array}{*{20}{c}}
   { - \left( {{x_s} + \frac{{{w_m}}}
{2} + (j - 1){w_m}} \right)}  \\
   {{y_s}}  \\
   { - \left( {{z_s} + \frac{{{w_m}}}
{2} + (i - 1){h_m}} \right)}  \\

 \end{array} } \right], \hfill \\
  {{\mathbf{P}}_k} = \left[ {\begin{array}{*{20}{c}}
   {{x_k} - \left( {{x_s} + \frac{{{w_m}}}
{2} + (j - 1){w_m}} \right)}  \\
   {{y_k}}  \\
   {{h_d} - \left( {{z_s} + \frac{{{w_m}}}
{2} + (i - 1){h_m}} \right)}  \\

 \end{array} } \right], \hfill \\
  {\mathbf{I}} = \left[ {\begin{array}{*{20}{c}}
   {{\mathbf{e}}_1^T\left( {{{\mathbf{R}}_{i,j}} + \frac{{{\mathbf{e}}_3^T({\mathbf{S}}-{\mathbf{R}})}}
{{{\mathbf{e}}_3^T\widehat{{{\mathbf{R}}_{i,j}}{\mathbf{I}}}}}\widehat{{{\mathbf{R}}_{i,j}}{\mathbf{I}}}} \right)}  \\
   {{\mathbf{e}}_2^T\left( {{{\mathbf{R}}_{i,j}} + \frac{{{\mathbf{e}}_3^T({\mathbf{S}}-{\mathbf{R}})}}
{{{\mathbf{e}}_3^T\widehat{{{\mathbf{R}}_{i,j}}{\mathbf{I}}}}}\widehat{{{\mathbf{R}}_{i,j}}{\mathbf{I}}}} \right)}  \\
   {{\mathbf{e}}_3^T{\mathbf{S}}}  \\

 \end{array} } \right], \hfill \\ 
\end{gathered}
\end{equation} 
where $\mathbb{R}_{i,j}^{r}$ represents the rotation matrix between the coordinate system in Fig.2a and Fig.2b, $x_s$, $y_s$ and $z_s$ represent the distance between the center of AP and the top left corner of the mirror array on x-axis y-axis and z-axis, respectively, $x_k$, $y_k$ and $h_d$ represent the distance between the center of AP and the position of user $k$ on x-axis y-axis and z-axis, respectively, as shown in Fig. 1. 

In (\ref{eq-IRS-irradience}), the binary indicator function $\mathbb{I}(\cdot)$  is used to indicate that the pre-reflection image of ${\bf{P}}_k$ (i.e. the point ${\bf{I}}$) locates within the boundaries of optical AP. 
In addition, ${\widehat{\mathbf{N}}_{i,j}}$ denotes the normal vector of the surface of the mirror ($i,j$) after rotation. According to Fig. 2, ${\widehat{\mathbf{N}}_{i,j}}$ can be expressed as  
\begin{equation}
{\widehat{\mathbf{N}}_{i,j}} = {\text{ }}\left[ {\begin{array}{*{20}{c}}
   {\sin ({\beta _{i,j}})\cos ({\alpha _{i,j}})}  \\
   {\cos ({\beta _{i,j}})\cos ({\alpha _{i,j}})}  \\
   {\sin ({\alpha _{i,j}})}  \\

 \end{array} } \right].
\end{equation}
Furthermore, $\widehat{{{\mathbf{R}}_{i,j}}{\mathbf{I}}}$ represents a unit vector along ${{\mathbf{R}}_{i,j}}{\mathbf{I}}$, which can be calculated by 
\begin{equation}
\widehat{{{\mathbf{R}}_{i,j}}{\mathbf{I}}} = 2\cos (\theta '){\widehat{\mathbf{N}}_{i,j}} - {{({{\mathbf{P}}_k} - {{\mathbf{R}}_{i,j}})} \mathord{\left/
 {\vphantom {{({{\mathbf{P}}_k} - {{\mathbf{R}}_{i,j}})} {{{\left\| {({{\mathbf{P}}_k} - {{\mathbf{R}}_{i,j}})} \right\|}_2}}}} \right.
 \kern-\nulldelimiterspace} {{{\left\| {({{\mathbf{P}}_k} - {{\mathbf{R}}_{i,j}})} \right\|}_2}}},
\end{equation}
where $\theta '$ represents the angle between the vector ${\widehat{\mathbf{N}}_{i,j}}$ and the reflected light from ${{\mathbf{R}}_{i,j}}$ to ${{{\mathbf{P}}_{k}}}$, which can be calculated by $\cos (\theta ') = {{\widehat{\mathbf{N}}_{i,j}^T({{\mathbf{P}}_k} - {{\mathbf{R}}_{i,j}})} \mathord{\left/
 {\vphantom {{\widehat{\mathbf{N}}_{i,j}^T({{\mathbf{P}}_k} - {{\mathbf{R}}_{i,j}})} {{{\left\| {{{\mathbf{P}}_k} - {{\mathbf{R}}_{i,j}}} \right\|}_2}}}} \right.
 \kern-\nulldelimiterspace} {{{\left\| {{{\mathbf{P}}_k} - {{\mathbf{R}}_{i,j}}} \right\|}_2}}}$. 
 Similarly, $\theta _{{{\mathbf{R}}_{i,j}}}^{\mathbf{I}}$ in (\ref{eq-IRS-irradience}) represents the angle between the vector ${\mathbf{e}}_3$ and the incident light from ${\mathbf{I}}$ to ${{{\mathbf{R}}_{i,j}}}$, which can be calculated by 
 $\theta _{{{\mathbf{R}}_{i,j}}}^{\mathbf{I}} = {\mathbf{e}}_3^T{{{({{\mathbf{R}}_{i,j}} - {\mathbf{I}})} \mathord{\left/
 {\vphantom {{({{\mathbf{R}}_{i,j}} - {\mathbf{I}})} {\left\| {{{\mathbf{R}}_{i,j}} - {\mathbf{I}}} \right\|}}} \right.
 \kern-\nulldelimiterspace} {\left\| {{{\mathbf{R}}_{i,j}} - {\mathbf{I}}} \right\|}}_2}$. In addition, $\theta _{{\mathbf{R}}_{i,j}}^{{\mathbf{P}}_k}$ in (\ref{eq-IRS-gain}) represents the the angle between the vector ${\mathbf{e}}_3$ and the reflected light from ${{{\mathbf{R}}_{i,j}}}$ to ${{\mathbf{P}}_k}$, which can be calculated by  $\theta _{{{\mathbf{R}}_{i,j}}}^{{\mathbf{P}}_k} = {\mathbf{e}}_3^T{{{({{\mathbf{R}}_{i,j}} - {\mathbf{I}})} \mathord{\left/ {\vphantom {{({{\mathbf{R}}_{i,j}} - {\mathbf{I}})} {\left\| {{{\mathbf{R}}_{i,j}} - {{\mathbf{P}}_k}} \right\|}}} \right. \kern-\nulldelimiterspace} {\left\| {{{\mathbf{R}}_{i,j}} - {{\mathbf{P}}_k}} \right\|}}_2}$.

\subsection{Secrecy rate maximization problem}
\label{section-OP}

Considering the peak power constraint of the optical transmitted signal, a lower bound of the achievable secrecy rate is given by \cite{Lampe-2015-JSAC}
\begin{equation}
\label{eq-Rsec}
\begin{array}{l}
R^{\rm{sec}}({\bm{\alpha }},{\bm{\beta }})
=\frac{1}{2}\log 
 \vspace{1ex} \\
\frac{{6{A^2}{{\left( {h^{{\rm{IRS}},\:\rm{Bob}}({\bm{\alpha }},{\bm{\beta }}) + {h}^{{\rm{LoS}},\:\rm{Bob}}} \right)}^T}\left( {h^{{\rm{IRS}},\:\rm{Bob}}({\bm{\alpha }},{\bm{\beta }}) + {h}^{{\rm{LoS}},\:\rm{Bob}}} \right) + 3\pi e{\sigma ^2}}}{{\pi e{A^2}{{\left( {h^{{\rm{IRS}},\:\rm{Eve}}({\bm{\alpha }},{\bm{\beta }}) + {h}^{{\rm{LoS}},\:\rm{Eve}}} \right)}^T}\left( {h^{{\rm{IRS}},\:\rm{Eve}}({\bm{\alpha }},{\bm{\beta }}) + {h}^{{\rm{LoS}},\:\rm{Eve}}} \right) + 3\pi e{\sigma ^2}}}.
\end{array}
\end{equation} 

In this paper, we propose a novel PLS technique for the VLC system with the help of IRS.
The key of the proposed PLS technique is how to intelligently control the orientation of each mirror to enhance the IRS channel gain of Bob while restricting the IRS channel gain of Eve simultaneously. 
Therefore, the ratio in (\ref{eq-Rsec}) can be enlarged by IRS, which results in the improvement of the achievable secrecy rate.
To maximize this improvement, we formulate the proposed IRS aided PLS into an achievable secrecy rate maximization problem, which is expressed as
\begin{subequations}
\label{Eq-OP-origin}
\begin{alignat}{2}
&\!\max_{\bm{\alpha},{{\bm{\beta}}}} &\qquad &  R^{\rm{sec}}({\bm{\alpha }},{\bm{\beta }})\label{Eq-OP-SISO-origin-objective}\\
&\text{subject to}\;(\rm{s.t.}) &      & - \frac{\pi }{2} \le {\alpha _{i,j}} \le \frac{\pi }{2}, \label{Eq-OP-SISO-origin-st-aerfa}\\
&                  &      & - \frac{\pi }{2} \le {\beta _{i,j}} \le \frac{\pi }{2}, \label{Eq-OP-SISO-PRF-st-beta}
\end{alignat}
\end{subequations}
where the orientation matrices of the mirror array are treated as optimization variables. 

For an IRS VLC system with $N_m \times N_n$ mirrors, the problem (\ref{Eq-OP-origin}) is a non-convex problem with $2 \times N_m \times N_n$ optimization variables.
Since the reflected spot of the optical AP in the receive plane is sensitive to the orientation of the mirror, the IRS channel gain of the mirror ($i,j$) is sensitive to both ${\alpha _{i,j}}$ and ${\beta _{i,j}}$. 
Because the aim of the problem (\ref{Eq-OP-origin}) is to find the optimal combination of ${\alpha _{i,j}}$ and ${\beta _{i,j}}$ for each mirror and find the optimal combination of orientations of all mirrors in the mirror array, the sensitivity of the variables on the objective function makes the high dimensional problem (\ref{Eq-OP-origin}) more intractable.  

Considering the combination of ${\alpha _{i,j}}$ and ${\beta _{i,j}}$ has one-to-one correspondence to the position of the reflected spot in the receive plane, we propose a reflected spot finding (RSF) method to reduce the dimension of the optimization variables in problem (\ref{Eq-OP-origin}). 
Instead of optimizing two angles of orientation of each mirror, we try to find the optimal position of reflected spot in the receive plane. 

In the local Cartesian coordinate system with the origin at the mirror center ${{\mathbf{R}}_{i,j}}$, the space coordinates of the center position of the reflected spot can be expressed as
\begin{equation}
\label{eq-Q}
{{\mathbf{Q}}} = \left[ {\begin{array}{*{20}{c}}
   {{x_q} - \left( {{x_s} + \frac{{{w_m}}}
{2} + (j - 1){w_m}} \right)}  \\
   {{y_q}}  \\
   {{h_q} - \left( {{z_s} + \frac{{{w_m}}}
{2} + (i - 1){h_m}} \right)}  \\

 \end{array} } \right], \hfill
\end{equation}
where $x_q$, $y_q$ and $h_q$ represent the distance between the center of AP and the center position of the reflected spot on x-axis y-axis and z-axis, respectively. 
According to the Snell's law of reflection, the reflected ray, the incident ray, and the normal vector of the surface of the mirror all lie in the same plane. In addition, the angle between the incident ray and the normal vector is equal to the angle between the reflected ray and the normal vector.
Therefore, for a target center position of reflected spot ${\mathbf{Q}}$, the normal vector of the surface of the mirror ($i,j$) after rotation can be calculated as 
\begin{equation}
\label{eq-N-snell}
{{{\mathbf{\hat N}}}_{i,j}} = \frac{{{{\mathbf{N}}_{i,j}}}}
{{{{\left\| {{{\mathbf{N}}_{i,j}}} \right\|}_2}}},
\end{equation}
where
\begin{equation}
{{\mathbf{N}}_{i,j}} = \frac{{{\mathbf{S}} - {{\mathbf{R}}_{i,j}}}}
{{{{\left\| {{\mathbf{S}} - {{\mathbf{R}}_{i,j}}} \right\|}_2}}} + \frac{{{\mathbf{Q}} - {{\mathbf{R}}_{i,j}}}}
{{{{\left\| {{\mathbf{Q}} - {{\mathbf{R}}_{i,j}}} \right\|}_2}}}.
\end{equation}
Furthermore, based on the obtained normal vector, the rotation angles can be calculated as 
\begin{equation}
\label{eq-angles-by-N}
\left\{ \begin{gathered}
  {\beta _{i,j}} = {\sin ^{ - 1}}\left( {{\mathbf{\hat N}}_{i,j}^T{{\mathbf{e}}_3}} \right) \hfill \\
  {\alpha _{i,j}} = {\sin ^{ - 1}}\left( {\frac{{{\mathbf{\hat N}}_{i,j}^T{{\mathbf{e}}_1}}}
{{\cos \left( {{\beta _{i,j}}} \right)}}} \right) \hfill \\ 
\end{gathered}  \right.
\end{equation}

By substituting (\ref{eq-N-snell}) into (\ref{eq-IRS-irradience}) and then substituting (\ref{eq-IRS-irradience}) into (\ref{eq-IRS-gain}) and (\ref{eq-Rsec}), we can obtain the achievable secrecy rate as a function of the center position of the reflected spot ${{\mathbf{Q}}}$, i.e. $R^{\rm{sec}}({\bf{Q}})$. 
Further, treating the function $R^{\rm{sec}}({\bf{Q}})$ as a new objective function, we transform the problem (\ref{Eq-OP-origin}) into
\begin{subequations}
\label{Eq-OP-PRF} 
\begin{alignat}{2}
&\!\max_{x_q,y_q,h_q} &\qquad & R^{\rm{sec}}({\bf{Q}}) \label{Eq-OP-PRF-objective}\\
&\rm{s.t.} &      & { {-x_0+x_s}\le {x_q}  \le {x_r-x_0+x_s} }, \label{Eq-OP-PRF-st-x}\\
&          &      & {0 \le y_q \le y_r}, \label{Eq-OP-PRF-st-y}\\
&          &      & {h_q=h_d}, \label{Eq-OP-PRF-st-z}
\end{alignat}
\end{subequations}
where $x_0$ denotes the distance between the left edge of the room and the left top corner of the mirror array on the x-axis, as shown in Fig. 1. The constraints (\ref{Eq-OP-PRF-st-x})-(\ref{Eq-OP-PRF-st-z}) are used to limit the reflected spot locates within the boundaries of the receive plane. 

Since the maximum IRS channel gain is achieved when all the mirrors focus on the same reflected spot, we just need to find one position of reflected spot for the mirror array in problem (\ref{Eq-OP-PRF}). In other words, just $x_q$, $y_q$ and $h_q$ need to be optimized and then the coordinates of $\mathbf{Q}$ for each local Cartesian coordinate system can be obtained. 
Therefore, compared to problem (\ref{Eq-OP-origin}), the dimension of optimization variables in the transformed problem by adopting the proposed RSF method is reduced from $2 \times N_m \times N_n$ to $3$.
\subsection{The modified PSO algorithm}
In this paper, the particle swarm optimization-initialization intervention (PSO-II) algorithm is leveraged to solve the non-convex problem (\ref{Eq-OP-PRF}) \cite{Qian}. The PSO is a heuristic algorithm which imitates swarms behaviour in birds flocking and fish schooling \cite{PSO1999}. 
In this paper, the position of each particle represents a potential solution of the problem (\ref{Eq-OP-PRF}), which is a three-dimensional variable including $x_q$, $y_q$ and $h_q$. 
The fitness of each particle is defined to measure the optimality of potential solutions, which is calculated by the objective function (\ref{Eq-OP-PRF-objective}).
Each particle searches for the maximum fitness by iteratively updating its position and velocity as follows
\begin{equation}
\label{equ-PSO-velocity}
\begin{aligned}
\bm{V}_{\lambda}^{n + 1} = &\zeta\bm{V}_{\lambda}^{n} + {c_1}{r_1}(\bm{W} _{\lambda,pbest}^{n} - \bm{W} _\lambda ^{n})\\
&+ {c_2}{r_2}(\bm{W}_{gbest}^{n} - \bm{W}_\lambda ^{n}), 
\end{aligned}
\end{equation}
\begin{equation}
\label{equ-PSO-position}
\bm{W}_{\lambda}^{n + 1} = \bm{W} _\lambda^{n}+\bm{V}_\lambda^{n+1}, 
\end{equation}
where $\bm{V}_{\lambda}^{n}$ and $\bm{W} _\lambda ^{n}$ are the velocity and position of particle $\lambda$ at the $n$th iteration, respectively. Furthermore, $\bm{W} _{\lambda,pbest}^{n}$ and $\bm{P}_{gbest}^{n}$ are the best position record of particle $\lambda$ and the entire swarm, respectively. In addition, $\zeta$ is the inertia weight, $c_1$ and $c_2$ are the learning factors, $r_1$ and $r_2$ are random numbers uniformly distributed in the range $\left[ {0,1} \right]$.

Since the non-convex objective function is sensitive to the position of the reflected spot, the convergence of PSO is poor if the positions of particles are initialized purely randomly as the conventional PSO. 
Therefore, we adopt PSO-II to improve the convergence performance, in which we intervene the initialization by choosing the position of Bob as one of potential solution of problem (\ref{Eq-OP-PRF}). 
The pseudocode of the PSO-II algorithm is shown in Algorithm 1.
\begin{table}
\begin{center}
\resizebox{0.5\textwidth}{!}{
\begin{tabular}{l}
\hline\noalign{\smallskip}
\textbf{Algorithm 1:} The PSO-II Algorithm\\
\noalign{\smallskip}\hline
\textbf{Input:} Swarm size $\lambda_{\rm{max}}$, maximum iterations $n_{\rm{max}}$; \\
\textbf{Initialization:} \\
\quad Position of the $1$st particle: ${{\bm{W}}_1}= \left[x_{\rm{Bob}}, y_{\rm{Bob}},h_{d} \right]$;\\
\quad Position of the other particles:  $\bm{W}_\lambda= \left[x_{q,\lambda}, y_{q,\lambda},h_{d} \right], 2 \le \lambda \le \lambda _{\rm{max}} $;\\
\quad Velocity of each particle: ${\bm{V}_\lambda }, 1 \le \lambda \le \lambda _{\rm{max}}$;\\
\quad Particle's best known position: 
${\bm{W}}_{\lambda,pbest}= \left[x_{q,\lambda,pbest}, y_{q,\lambda,pbest},h_{d} \right], $\\
\quad \quad \quad \quad \quad \quad \quad \quad \quad \quad \quad \quad \quad \quad $1 \le \lambda \le \lambda _{\rm{max}}$; \\
\quad Swarm's best known position: ${\bm{W}}_{\lambda,gbest}= \left[x_{q,\lambda,gbest}, y_{q,\lambda,gbest},h_{d} \right]$; \\
\textbf{While} $n \le n_{\rm{max}}$\\
\quad \textbf{For} each particle \\
\quad \quad Update the velocity and the position according to (\ref{equ-PSO-velocity}) and (\ref{equ-PSO-position}); \\
\quad \quad Calculate the fitness of each particle by (\ref{Eq-OP-PRF-objective});\\
\quad \quad \textbf{If} the fitness value of ${\bm{W}}_\lambda^{n}$ is lager than the fitness value of ${\bm{W}}_{\lambda,pbest}^{n-1}$ \\
\quad \quad \quad Update the particle's best known position as ${\bm{W}}_{\lambda,pbest}^{n}={\bm{W}}_\lambda^{n}$;\\
\quad \quad \textbf{End if} \\
\quad \quad \textbf{If} the fitness value of ${\bm{W}}_\lambda^{n}$ is smaller than the fitness value of ${\bm{W}}_{gbest}^{n-1}$\\
\quad \quad \quad Update the swarm's best known position as ${\bm{W}}_{gbest}^{n}={\bm{W}}_\lambda^{n}$; \\
\quad \quad \textbf{End if} \\
\quad \textbf{End for}\\
\textbf{End while}\\
\textbf{Output:} ${x_q^{*}}$,${y_q^{*}}$ and ${h_q^{*}}$ in the swarm's best known position ${\bm{W}}_{gbest}^{n_{\rm{max}}}$. \\
\noalign{\smallskip}\hline
\end{tabular}}
\end{center}
\vspace*{-4mm}
\end{table}

\section{Simulations Results and Discussion}
\label{section-simulation}
In this section, we present the simulation results to evaluate the secrecy performance of the IRS aided VLC system. The main system parameters are summarized in Table I.
We calculate the integrals in (\ref{eq-LoS-gain}) and (\ref{eq-IRS-irradience}) by a discretization method, and the area of each discrete element is $10^{-4} \times 10^{-4} \; {\rm{m}}^2$.

\begin{table}[htbp]
\caption{Simulation parameters}
\begin{center}
\begin{tabular}{ll}      
\hline\noalign{\smallskip}
\textbf{Name of Parameter} & \textbf{Value of Parameter}   \\
\noalign{\smallskip}\hline\noalign{\smallskip}
Room size, $x_r \times y_r \times z_r$ & $5 \times 5 \times 3\;\rm{m}^3$ \\
AP size, $w_s \times l_s$ & $0.01 \times 0.01\;\rm{m}^2$\\
The position of AP, $(x_s,y_s,z_s)$ & (-0.26,  2.5, 0.5) m\\
The position of Bob, $(x_{\rm{Bob}},y_{\rm{Bob}},h_d)$ & (0.2, 2, 3) m\\
The position of mirror array, $x_0$ & 2.24 m\\
Peak power constraint, $A$ & 140 mA\\
Refractive index of lens at a PD, $a$  & 1.5  \\
Physical area in a PD, $\delta$   & $10^{-4} \;\rm{m}^2$\\
Semi-angle at half power, ${\phi _{\;1/2}}$ & 70 deg. \\
Gain of the transimpedance amplifier, $T$  & 1 V/A\\
Current-to-light conversion efficiency, $\eta$ & 0.44 W/A\\
Responsivity of a PD, $\varpi$ & 0.54 A/W\\
Reflection efficiency, $\rho$ & 0.8 \\
\noalign{\smallskip}\hline
\end{tabular}
\label{Table-parameters}
\end{center}
\vspace*{-5mm}
\end{table}


To show the improvement of secrecy performance by adding IRS in the VLC system, we present Fig. \ref{Fig-channel-gain} and Fig. \ref{Fig-Csec}, in which we compare the following three mirror orientation control methods.
\begin{itemize}
\item The proposed RSF method: We solve the problem (\ref{Eq-OP-PRF}) to obtain the optimal center position of the reflected spot and then determine rotation angles of mirrors by formula (\ref{eq-angles-by-N}).
\item Focus on Bob (FoB) method: We choose Bob's position as the center position of the reflected spot, and then determine rotation angles of mirrors by formula (\ref{eq-angles-by-N}).
\item Without IRS (w/o IRS): Only LoS links are considered in the system without IRS.
\end{itemize}  

Fig. \ref{Fig-channel-gain} and Fig. \ref{Fig-Csec} show the channel gain and the achievable secrecy rate for various positions of Eve, respectively, where the IRS contains $5 \times 5$ mirrors and the size of each mirror is $0.1 \times 0.1 \rm{m}^2$. The AP is installed in the center of the ceiling. 
Bob and Eve have the same coordinates on y-axis and z-axis, while on x-axis, the distance between AP and Bob (i.e. $x_{\rm{Bob}}$) is 0.2 and the distance between AP and Eve (i.e. $x_{\rm{Eve}}$) changes from -1 to 1. 

When there is no IRS in the VLC system, only LoS channel gains are considered. 
As shown in Fig. \ref{Fig-channel-gain}, Eve's LoS channel gain first increases then decreases when Eve first comes close to AP then goes away from AP.
In this case, since the difference of the channel gain between Bob and Eve is relatively small, the corresponding achievable secrecy rate is limited and equals to zero for $-0.8<x_{\rm{Eve}}<0.8$, which means the insecure area is large, as shown in Fig. \ref{Fig-Csec}.

When we add IRS in the VLC system and using FoB method to control mirrors orientations, the IRS gain of Bob is positive while the IRS gain of Eve is zero for most cases except $0.1<x_{\rm{Eve}}<0.3$. 
This is because we consider a light source with an area, which cannot be described as point source, and the reflected spot in the receive plane is also an area. 
When Eve is close to Bob, Eve can also receive the optical signal reflected by IRS. Therefore, the difference between the sum channel gain of Bob and Eve is enlarged by IRS in most of cases except $0.1<x_{\rm{Eve}}<0.3$ and the corresponding achievable secrecy rate is improved, as shown in Fig. \ref{Fig-channel-gain} and Fig. \ref{Fig-Csec}, respectively.

Although the FoB method enhances the secrecy performance with low complexity, the insecure area still exists. 
If we optimize the center position of the reflected spot according to the proposed RSF method, the insecure area can be further reduced.
As shown in Fig. \ref{Fig-Csec}, except the position of Bob and Eve are overlapped, the achievable secrecy rate is positive for all the cases. 
In other words, by controlling the mirrors' orientation more intelligently, Bob can always receive the reflected signal while Eve cannot. This is because by choosing an optimized center position of the reflected spot, we can make Bob still locate in the area of the reflected spot while Eve is removed out of the area of the reflected spot.

\definecolor{mycolor1}{rgb}{0.87059,0.49020,0.00000}%
\definecolor{mycolor2}{rgb}{0.00000,0.49804,0.00000}%
\definecolor{mycolor3}{rgb}{1.00000,0.00000,1.00000}%
\definecolor{mycolor4}{rgb}{0.00000,1.00000,1.00000}%
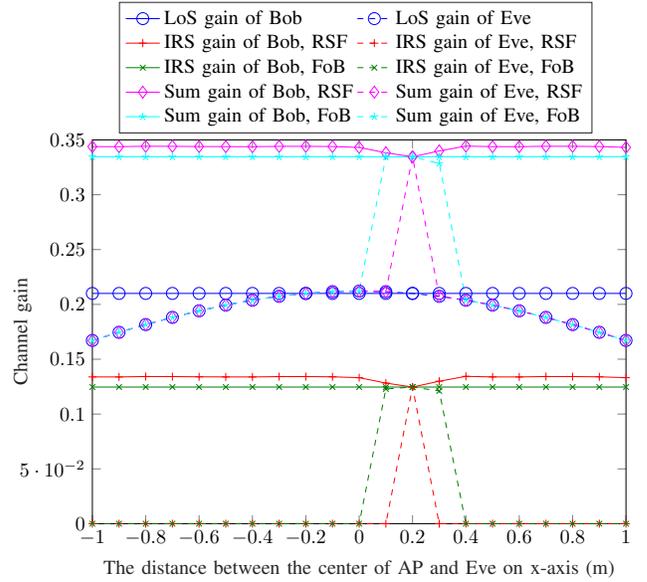
\begin{figure}[!t]
\centering
\resizebox{3.3in}{!}{%
\begin{tikzpicture}

\begin{axis}[%
width=3.708in,
height=2.669in,
at={(0.622in,0.447in)},
scale only axis,
xmin=-1,
xmax=1,
xlabel style={font=\color{white!15!black}},
xlabel={The distance between the center of AP and Eve on x-axis (m)},
ymin=0,
ymax=0.35,
ylabel style={font=\color{white!15!black}},
ylabel={Channel gain},
axis background/.style={fill=white},
legend style={at={(axis cs:-0.9,0.356)}, anchor=south west, legend columns=2, legend cell align=left, align=left, draw=white!15!black}
]
\addplot [color=blue, mark=o,  mark size=3.0pt, mark options={solid, blue}]
  table[row sep=crcr]{%
-1	0.209995657404435\\
-0.9	0.209995657404435\\
-0.8	0.209995657404435\\
-0.7	0.209995657404435\\
-0.6	0.209995657404435\\
-0.5	0.209995657404435\\
-0.4	0.209995657404435\\
-0.3	0.209995657404435\\
-0.2	0.209995657404435\\
-0.1	0.209995657404435\\
0	0.209995657404435\\
0.1	0.209995657404435\\
0.2	0.209995657404435\\
0.3	0.209995657404435\\
0.4	0.209995657404435\\
0.5	0.209995657404435\\
0.6	0.209995657404435\\
0.7	0.209995657404435\\
0.8	0.209995657404435\\
0.9	0.209995657404435\\
1	0.209995657404435\\
};
\addlegendentry{LoS gain of Bob}

\addplot [color=blue, dashed, mark=o, mark size=3.0pt, mark options={solid, blue}]
  table[row sep=crcr]{%
-1	0.167108549538856\\
-0.9	0.174532065352172\\
-0.8	0.181580607102304\\
-0.7	0.188143010371494\\
-0.6	0.19410834077006\\
-0.5	0.199369517998889\\
-0.4	0.203827159602599\\
-0.3	0.207393430426236\\
-0.2	0.209995657404435\\
-0.1	0.211579463950729\\
0	0.212111197120831\\
0.1	0.211579463950729\\
0.2	0.209995657404435\\
0.3	0.207393430426236\\
0.4	0.203827159602599\\
0.5	0.199369517998889\\
0.6	0.19410834077006\\
0.7	0.188143010371494\\
0.8	0.181580607102304\\
0.9	0.174532065352172\\
1	0.167108549538856\\
};
\addlegendentry{LoS gain of Eve}

\addplot [color=red, mark=+, mark options={solid, red}]
  table[row sep=crcr]{%
-1	0.133849900368258\\
-0.9	0.133824467569904\\
-0.8	0.13420840338793\\
-0.7	0.134156499681843\\
-0.6	0.133944183780738\\
-0.5	0.133849900368258\\
-0.4	0.133824467569904\\
-0.3	0.134114008905717\\
-0.2	0.134156499681843\\
-0.1	0.133944183780738\\
0	0.133158789673982\\
0.1	0.128257224256646\\
0.2	0.124613407645021\\
0.3	0.129930190328159\\
0.4	0.134365123500506\\
0.5	0.133849900368258\\
0.6	0.133824467569904\\
0.7	0.13420840338793\\
0.8	0.134156499681843\\
0.9	0.133944183780738\\
1	0.133158789673982\\
};
\addlegendentry{IRS gain of Bob, RSF}

\addplot [color=red, dashed, mark=+, mark options={solid, red}]
  table[row sep=crcr]{%
-1	0\\
-0.9	0\\
-0.8	0\\
-0.7	0\\
-0.6	0\\
-0.5	0\\
-0.4	0\\
-0.3	0\\
-0.2	0\\
-0.1	0\\
0	0\\
0.1	0\\
0.2	0.124613407645021\\
0.3	0\\
0.4	0\\
0.5	0\\
0.6	0\\
0.7	0\\
0.8	0\\
0.9	0\\
1	0\\
};
\addlegendentry{IRS gain of Eve, RSF}

\addplot [color=mycolor2, mark=x, mark options={solid, mycolor2}]
  table[row sep=crcr]{%
-1	0.124613407645021\\
-0.9	0.124613407645021\\
-0.8	0.124613407645021\\
-0.7	0.124613407645021\\
-0.6	0.124613407645021\\
-0.5	0.124613407645021\\
-0.4	0.124613407645021\\
-0.3	0.124613407645021\\
-0.2	0.124613407645021\\
-0.1	0.124613407645021\\
0	0.124613407645021\\
0.1	0.124613407645021\\
0.2	0.124613407645021\\
0.3	0.124613407645021\\
0.4	0.124613407645021\\
0.5	0.124613407645021\\
0.6	0.124613407645021\\
0.7	0.124613407645021\\
0.8	0.124613407645021\\
0.9	0.124613407645021\\
1	0.124613407645021\\
};
\addlegendentry{IRS gain of Bob, FoB}

\addplot [color=mycolor2, dashed, mark=x, mark options={solid, mycolor2}]
  table[row sep=crcr]{%
-1	0\\
-0.9	0\\
-0.8	0\\
-0.7	0\\
-0.6	0\\
-0.5	0\\
-0.4	0\\
-0.3	0\\
-0.2	0\\
-0.1	0\\
0	0\\
0.1	0.122875462234077\\
0.2	0.124613407645021\\
0.3	0.121089828413045\\
0.4	0\\
0.5	0\\
0.6	0\\
0.7	0\\
0.8	0\\
0.9	0\\
1	0\\
};
\addlegendentry{IRS gain of Eve, FoB}

\addplot [color=mycolor3, mark=diamond, mark size=3.0pt, mark options={solid, mycolor3}]
  table[row sep=crcr]{%
-1	0.343845557772693\\
-0.9	0.343820124974339\\
-0.8	0.344204060792365\\
-0.7	0.344152157086278\\
-0.6	0.343939841185173\\
-0.5	0.343845557772693\\
-0.4	0.343820124974339\\
-0.3	0.344109666310152\\
-0.2	0.344152157086278\\
-0.1	0.343939841185173\\
0	0.343154447078417\\
0.1	0.338252881661081\\
0.2	0.334609065049456\\
0.3	0.339925847732594\\
0.4	0.344360780904941\\
0.5	0.343845557772693\\
0.6	0.343820124974339\\
0.7	0.344204060792365\\
0.8	0.344152157086278\\
0.9	0.343939841185173\\
1	0.343154447078417\\
};
\addlegendentry{Sum gain of Bob, RSF}

\addplot [color=mycolor3, dashed, mark=diamond, mark size=3.0pt, mark options={solid, mycolor3}]
  table[row sep=crcr]{%
-1	0.167108549538856\\
-0.9	0.174532065352172\\
-0.8	0.181580607102304\\
-0.7	0.188143010371494\\
-0.6	0.19410834077006\\
-0.5	0.199369517998889\\
-0.4	0.203827159602599\\
-0.3	0.207393430426236\\
-0.2	0.209995657404435\\
-0.1	0.211579463950729\\
0	0.212111197120831\\
0.1	0.211579463950729\\
0.2	0.334609065049456\\
0.3	0.207393430426236\\
0.4	0.203827159602599\\
0.5	0.199369517998889\\
0.6	0.19410834077006\\
0.7	0.188143010371494\\
0.8	0.181580607102304\\
0.9	0.174532065352172\\
1	0.167108549538856\\
};
\addlegendentry{Sum gain of Eve, RSF}

\addplot [color=mycolor4, mark=star, mark options={solid, mycolor4}]
  table[row sep=crcr]{%
-1	0.334609065049456\\
-0.9	0.334609065049456\\
-0.8	0.334609065049456\\
-0.7	0.334609065049456\\
-0.6	0.334609065049456\\
-0.5	0.334609065049456\\
-0.4	0.334609065049456\\
-0.3	0.334609065049456\\
-0.2	0.334609065049456\\
-0.1	0.334609065049456\\
0	0.334609065049456\\
0.1	0.334609065049456\\
0.2	0.334609065049456\\
0.3	0.334609065049456\\
0.4	0.334609065049456\\
0.5	0.334609065049456\\
0.6	0.334609065049456\\
0.7	0.334609065049456\\
0.8	0.334609065049456\\
0.9	0.334609065049456\\
1	0.334609065049456\\
};
\addlegendentry{Sum gain of Bob, FoB}

\addplot [color=mycolor4, dashed, mark=star, mark options={solid, mycolor4}]
  table[row sep=crcr]{%
-1	0.167108549538856\\
-0.9	0.174532065352172\\
-0.8	0.181580607102304\\
-0.7	0.188143010371494\\
-0.6	0.19410834077006\\
-0.5	0.199369517998889\\
-0.4	0.203827159602599\\
-0.3	0.207393430426236\\
-0.2	0.209995657404435\\
-0.1	0.211579463950729\\
0	0.212111197120831\\
0.1	0.334454926184806\\
0.2	0.334609065049456\\
0.3	0.328483258839281\\
0.4	0.203827159602599\\
0.5	0.199369517998889\\
0.6	0.19410834077006\\
0.7	0.188143010371494\\
0.8	0.181580607102304\\
0.9	0.174532065352172\\
1	0.167108549538856\\
};
\addlegendentry{Sum gain of Eve, FoB}

\end{axis}
\end{tikzpicture}%
}
\vspace*{-2mm}
\caption{The channel gains of the IRS aided VLC system obtained by different mirror orientation control scheme. The IRS contains $5 \times 5$ mirrors and the size of each mirror is $0.1 \times 0.1 \rm{m}^2$.}
\label{Fig-channel-gain}
\vspace*{-6mm}
\end{figure}

\definecolor{mycolor1}{rgb}{0.74902,0.00000,0.74902}%
\definecolor{mycolor2}{rgb}{0.00000,0.49804,0.00000}%
\definecolor{mycolor3}{rgb}{0.92941,0.69412,0.12549}%

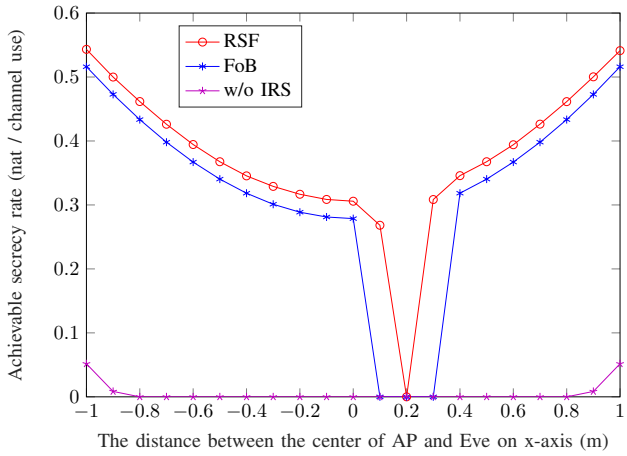
\begin{figure}[!t]
\centering
\resizebox{3.3in}{!}{%
\begin{tikzpicture}

\begin{axis}[%
width=3.708in,
height=2.669in,
at={(0.622in,0.447in)},
scale only axis,
xmin=-1,
xmax=1,
xlabel style={font=\color{white!15!black}},
xlabel={The distance between the center of AP and Eve on x-axis (m)},
ymin=0,
ymax=0.6,
ylabel style={font=\color{white!15!black}},
ylabel={Achievable secrecy rate (nat / channel use)},
axis background/.style={fill=white},
legend style={at={(0.173,0.754)}, anchor=south west, legend cell align=left, align=left, draw=white!15!black}
]
\addplot [color=red,mark=o, mark options={solid,red}]
  table[row sep=crcr]{%
-1	0.543251175389834\\
-0.9	0.499939442963088\\
-0.8	0.461652598302076\\
-0.7	0.426157760885028\\
-0.6	0.394457983776051\\
-0.5	0.367546176357525\\
-0.4	0.345442925016605\\
-0.3	0.329000492893704\\
-0.2	0.316698177232087\\
-0.1	0.308594199428498\\
0	0.305819061941522\\
0.1	0.268220611142291\\
0.2	0\\
0.3	0.308461865074352\\
0.4	0.345790418556026\\
0.5	0.367546176357525\\
0.6	0.39411049023663\\
0.7	0.426308288844042\\
0.8	0.461502070343061\\
0.9	0.500286936502509\\
1	0.541242915516691\\
};
\addlegendentry{RSF}

\addplot [color=blue,mark=asterisk, mark options={solid,blue}]
  table[row sep=crcr]{%
-1	0.516072926560884\\
-0.9	0.472835026576918\\
-0.8	0.433434179216596\\
-0.7	0.398089869758562\\
-0.6	0.36700607385046\\
-0.5	0.340367927528575\\
-0.4	0.318338508630436\\
-0.3	0.301055847760186\\
-0.2	0.288630286105621\\
-0.1	0.281142289502907\\
0	0.278640813112572\\
0.1	0\\
0.2	0\\
0.3	0\\
0.4	0.318338508630436\\
0.5	0.340367927528575\\
0.6	0.36700607385046\\
0.7	0.398089869758562\\
0.8	0.433434179216596\\
0.9	0.472835026576919\\
1	0.516072926560884\\
};
\addlegendentry{FoB}

\addplot [color=mycolor1,mark=star, mark options={solid,mycolor1}]
  table[row sep=crcr]{%
-1	0.0516889569436838\\
-0.9	0.00845105695971803\\
-0.8	0\\
-0.7	0\\
-0.6	0\\
-0.5	0\\
-0.4	0\\
-0.3	0\\
-0.2	0\\
-0.1	0\\
0	0\\
0.1	0\\
0.2	0\\
0.3	0\\
0.4	0\\
0.5	0\\
0.6	0\\
0.7	0\\
0.8	0\\
0.9	0.00845105695971825\\
1	0.0516889569436838\\
};
\addlegendentry{w/o IRS}

\end{axis}
\end{tikzpicture}%
}
\vspace*{-3mm}
\caption{The achievable secrecy rate obtained by different mirror orientation control scheme. The IRS contains $5 \times 5$ mirrors and the size of each mirror is $0.1 \times 0.1 \rm{m}^2$.}
\vspace*{-3mm}
\label{Fig-Csec}
\end{figure}

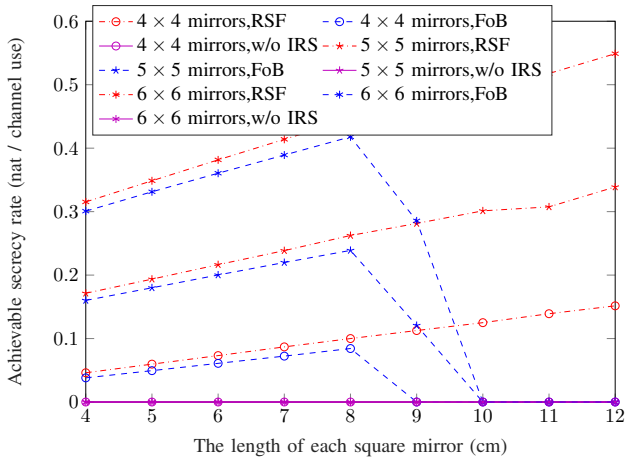
\begin{figure}[!t]
\centering
\resizebox{3.3in}{!}{%
\begin{tikzpicture}

\begin{axis}[%
width=3.708in,
height=2.669in,
at={(0.622in,0.447in)},
scale only axis,
xmin=4,
xmax=12,
xlabel style={font=\color{white!15!black}},
xlabel={The length of each square mirror (cm)},
ymin=0,
ymax=0.6,
ylabel style={font=\color{white!15!black}},
ylabel={Achievable secrecy rate (nat / channel use)},
axis background/.style={fill=white},
legend style={at={(axis cs:4.1,0.42)}, anchor=south west, legend columns=2, legend cell align=left, align=left, draw=white!15!black}
]
\addplot [color=red, dashdotted, mark=o, mark options={solid, red}]
  table[row sep=crcr]{%
4	0.0461055515397542\\
5	0.0596756932702154\\
6	0.0731410083123541\\
7	0.0868744354864461\\
8	0.0999779414131383\\
9	0.112636583600182\\
10	0.125043307412991\\
11	0.139150817851691\\
12	0.151575152179034\\
};
\addlegendentry{$\text{4}\times\text{4 mirrors,RSF}$}

\addplot [color=blue, dashed, mark=o, mark options={solid, blue}]
  table[row sep=crcr]{%
4	0.0380628202720402\\
5	0.0494338804818047\\
6	0.0609369474045944\\
7	0.0723283032077466\\
8	0.0842868820513399\\
9	0\\
10	0\\
11	0\\
12	0\\
};
\addlegendentry{$\text{4}\times\text{4 mirrors,FoB}$}

\addplot [color=mycolor1, mark=o, mark options={solid, mycolor1}]
  table[row sep=crcr]{%
4	0\\
5	0\\
6	0\\
7	0\\
8	0\\
9	0\\
10	0\\
11	0\\
12	0\\
};
\addlegendentry{$\text{4}\times\text{4 mirrors,w/o IRS}$}

\addplot [color=red, dashdotted, mark=star, mark options={solid, red}]
  table[row sep=crcr]{%
4	0.171381066591703\\
5	0.193601841351951\\
6	0.216448247351176\\
7	0.238574082933051\\
8	0.262414079784946\\
9	0.28140797221369\\
10	0.301261921710655\\
11	0.307492393029482\\
12	0.33888182038867\\
};
\addlegendentry{$\text{5}\times\text{5 mirrors,RSF}$}

\addplot [color=blue, dashed, mark=star, mark options={solid, blue}]
  table[row sep=crcr]{%
4	0.160103223906746\\
5	0.17997235852137\\
6	0.200201086603236\\
7	0.219856902498936\\
8	0.238901302187506\\
9	0.120629977842965\\
10	0\\
11	0\\
12	0\\
};
\addlegendentry{$\text{5}\times\text{5 mirrors,FoB}$}

\addplot [color=mycolor1, mark=star, mark options={solid, mycolor1}]
  table[row sep=crcr]{%
4	0\\
5	0\\
6	0\\
7	0\\
8	0\\
9	0\\
10	0\\
11	0\\
12	0\\
};
\addlegendentry{$\text{5}\times\text{5 mirrors,w/o IRS}$}

\addplot [color=red, dashdotted, mark=asterisk, mark options={solid, red}]
  table[row sep=crcr]{%
4	0.315328114433941\\
5	0.34861716406436\\
6	0.381574385124701\\
7	0.413909346901007\\
8	0.445713479533526\\
9	0.46325526600373\\
10	0.490275254763371\\
11	0.517659484711801\\
12	0.54885190398189\\
};
\addlegendentry{$\text{6}\times\text{6 mirrors,RSF}$}

\addplot [color=blue, dashed, mark=asterisk, mark options={solid, blue}]
  table[row sep=crcr]{%
4	0.30074765841582\\
5	0.331256001483715\\
6	0.360505177473407\\
7	0.389176129333446\\
8	0.417800518785727\\
9	0.285447022809717\\
10	0\\
11	0\\
12	0\\
};
\addlegendentry{$\text{6}\times\text{6 mirrors,FoB}$}

\addplot [color=mycolor1, mark=asterisk, mark options={solid, mycolor1}]
  table[row sep=crcr]{%
4	0\\
5	0\\
6	0\\
7	0\\
8	0\\
9	0\\
10	0\\
11	0\\
12	0\\
};
\addlegendentry{$\text{6}\times\text{6 mirrors,w/o IRS}$}
\end{axis}
\end{tikzpicture}%
}
\vspace*{-3mm}
\caption{The achievable secrecy rate provided by different mirror orientation control schemes for various mirror size. The distance between AP and Eve is 0.1.}
\vspace*{-4mm}
\label{Fig-mirr-size}
\end{figure}

Fig. \ref{Fig-mirr-size} shows the impact of the size of the mirror array and the size of each mirror on the achievable secrecy rate when $x_{\rm{eve}}=0.1$. Each mirror is a square.
Since the eavesdropping channel is stronger than the legitimate communication channel in this figure, the VLC system without IRS is insecure and the achievable secrecy rate is zero. 
For an IRS VLC system using the FoB method, when the mirror size is small, the achievable secrecy rate increases with the size of mirror increasing.
This is because a large size mirror can increase the effective receive area. 
When the area of the reflected spot is large enough and the center of the reflected spot still lies in Bob, Eve in the neighbourhood of Bob can also receive the reflected signal, which leads to the achievable secrecy rate reduces to zero.
In contrast, for the proposed RSF method, since the position of the reflected spot is optimized adaptively, the system can maintain secure all the time.

\section{Conclusion}
\label{section-conclusion}
In this paper, we proposed a novel PLS technique for VLC system with the help of IRS which is implemented by a mirror array. By intelligently controlling the orientation of each mirror, we enlarge the difference between the channel gain of Bob and Eve and improve the secrecy performance. 
This work is the first step to explore the IRS in the secure VLC system. In the future, IRS aided PLS can be extend to more scenarios, such as multiple APs, multiple users, etc.

\section*{Acknowledgment}
The research is supported in part by National Natural Science Foundation of China (No.61801191), Jilin Scientific and Technological Development Program (No.20180101040JC and No.20200401147GX), the China Scholarship Council under (No.201906170201), the Natural Sciences and Engineering Research Council of Canada (NSERC).


\ifCLASSOPTIONcaptionsoff
  \newpage
\fi



%



\bibliographystyle{IEEEtran}
\bibliography{IEEEabrv,references-Lei-paper-IRSVLC}

%

%
%
%
%
%
%
%
%
%
%
%




\end{document}